\documentclass[aip,amsmath,amssymb,reprint]{revtex4-1}

\usepackage{chemformula} % Formula subscripts using \ch{}
\usepackage[T1]{fontenc} % Use modern font encodings

\begin{document}
	\preprint{AIP/123-QED}

\title[]{Interaction of CH$_3$CN and CH$_3$NC with He : potential energy surfaces and low-energy scattering}

\author{Malek Ben Khalifa}
%\email{malek.benkhalifa@kuleuven.be}
\affiliation{KU Leuven, Department of Chemistry, Celestijnenlaan 200F, B-3001 Leuven, Belgium.}
\author{Paul J. Dagdigian}
\affiliation{The Johns Hopkins University, Department of Chemistry, Baltimore, MD 21218-2685, USA.}
\author{J\'er\^ome Loreau}
\affiliation{KU Leuven, Department of Chemistry, Celestijnenlaan 200F, B-3001 Leuven, Belgium.}
\email{jerome.loreau@kuleuven.be}

\begin{abstract}
  Several nitrogen-bearing molecules, such as methyl cyanide (or acetonitrile, CH$_3$CN) and methyl isocyanide (CH$_3$NC) of interest here, have been observed in various astrophysical environments. The accurate modeling of their
abundance requires the calculation of rate coefficients for their collisional excitation with species such as He atoms or H$_2$ molecules at low temperatures. In this work we compute new three-dimensional potential energy
surfaces for the CH$_3$NC-He and CH$_3$CN-He van der Waals complexes by means of the explicitly correlated coupled cluster approach with single, double and perturbative triple excitation CCSD(T)/F12a in conjunction with the aug-cc-pVTZ basis set. We find a global minimum with $D_e= 55.10$ and 58.61 cm$^{-1}$ for CH$_3$CN-He and CH$_3$NC-He, respectively, while the dissociation energy $D_0$ of the complexes are 18.64 and 18.65 cm$^{-1}$, respectively. 
Low energy scattering calculations of pure rotational (de-)excitation of CH$_3$CN and CH$_3$NC by collision with He atoms are carried out with the close-coupling method and the collisional cross sections of $ortho-$ and $para-$CH$_3$NC and
CH$_3$CN are computed for kinetic energies up to 100 cm$^{-1}$. While the PESs for both complexes are qualitatively similar, that of CH$_3$NC-He is more anisotropic, leading to different propensity rules for rotational excitation.
For CH$_3$NC-He, we find that |$\Delta j$| = 1 transitions are dominant at low kinetic energy and a propensity rule that favors odd $\Delta j$ transitions is observed, whereas for CH$_3$CN the dominant cross sections are associated to transitions with |$\Delta j$| = 2.
We expect that the findings of this study will be beneficial for astrophysical investigations as well as laboratory experiments.
\end{abstract}

	\maketitle

\section{Introduction}
Methyl cyanide (or acetonitrile, CH$_3$CN) is one of the most ubiquitous interstellar organic molecules. It has been detected in a variety of low- and high-mass sources \cite{beltran2005detailed,zapata2010rotating}. 
Thanks to its three-fold symmetry and its large dipole moment, CH$_3$CN is often used as a gas thermometer to probe interstellar environments and it is a useful tracer of dense gas \cite{wilner1994maps, Andron2018}. 
Its isomer CH$_3$NC is thermodynamically the less stable isomer of CH$_3$CN. CH$_3$NC was tentatively observed for the first time in 1988 in the Sgr B2 cloud \cite{cernicharo1988tentative} and confirmed with additional rotational transitions seventeen years later by \citet{remijan2005interstellar}
Since then, it has been observed only in a handful of sources \cite{calcutt2018alma}, including the Horsehead nebula photodissociation region (PDR) \cite{gratier2013iram}, as well as Orion KL \cite{lopez2014laboratory}.

For typical dense cloud conditions, theoretical calculations by \citet{defrees1985theoretical} estimated the ratio of CH$_3$NC/CH$_3$CN to be in the range of $0.1-0.4$. However, \citet{cernicharo1988tentative} deduced from their
observations an abundance ratio of $\sim 0.03-0.05$, showing that the relative abundance of CH$_3$CN is larger than previous estimates. Later, \citet{gratier2013iram} found, according to their observations in the Horsehead region, an abundance ratio CH$_3$NC/CH$_3$CN in agreement with estimations of Defrees. Therefore, it appears that accurately determining the abundances of CH$_3$CN and CH$_3$NC may provide clues to understand their respective chemistry in space, and the formation and evolution of interstellar clouds can be explored through isomer comparison studies.

Acetonitrile and methyl isocyanide are symmetric top molecules that display close similarities in their molecular parameters such as the bond length \cite{moffat1964lcao,costain1958determination} and electric dipole moment \cite{lee1972microwave}, which cannot introduce any bias in their respective rotational spectra. 
Unfortunately, the abundance of CH$_3$NC is not well constrained as its abundance is estimated using excitation rate coefficients of CH$_3$CN, an approximation that might not be valid. 
In order to satisfy astrophysical precision when modelling the abundance of interstellar species, we need calculations of rate coefficients for rotational excitation induced by collision with the most abundant species (H$_2$, He, and H). 

In the present paper, we study the collision of the rigid symmetric top molecules CH$_3$CN and CH$_3$NC with He at low collision energies ($E_{coll} \leq 100$ cm$^{-1}$), with the aim of understanding their excitation and the impact of isomerization on the rotational excitation. 
A PES for CH$_3$CN-He has been reported recently \cite{khalifa2020rotational} and used to investigate the rotational excitation of CH$_3$CN. However, due to an error in the angular expansion of the potential,
the radial coefficients used for the scattering computations were not correct.
As a consequence, the CH$_3$CN-He cross sections were impacted by an error that can reach an order of magnitude for some transitions, making a meaningful comparison with CH$_3$NC-He cross sections impossible.
Consequently, in order to establish a detailed comparison between the interaction and collisional excitation properties of methyl cyanide and methyl isocyanide with helium atoms, we compute new PESs for both isomers at exactly the same computational level and basis set.

Finally, we note that the reactivity of methyl cyanide with molecular ions at low temperature has been studied experimentally by several groups \cite{Okada2013,Okada2020,Krohn2021}. Ideally, this requires molecules that are cooled prior to the reaction. Several approaches to produce cold beams of CH$_3$CN have been reported \cite{Gandhi1987, Liu2010, Okada2013}, which allow the production of CH$_3$CN in the ground rotational state \cite{Spieler2013}. Another possibility would be to use a Stark decelerator. Cooling can also be achieved by loading the CH$_3$CN molecules into a helium buffer gas cell, and the knowledge of the rotational excitation cross sections and rate coefficients can be used to model the dynamics in the cell \cite{Schullian2015,Doppelbauer2017}. 

The overall structure of this paper is as follows: first, we present in Section \ref{sec:PES} an \textit{ab initio} study of the CH$_3$CN/NC-He systems, leading to new 3D-PESs corresponding to the interaction between CH$_3$CN/NC and He. Section \ref{sec:bound} is dedicated to the investigation of the dissociation energy of the complexes, while in Section \ref{sec:scat} we describe the study of the dynamics of the collision, restricted in the present article to the low-energy regime, which we illustrate with inelastic cross sections in CH$_3$CN-He and CH$_3$NC-He collisions.
Conclusions and future outlooks are drawn in Section \ref{sec:conclusion}.

\section{Potential energy surfaces}\label{sec:PES}
\subsection{\textit{Ab initio} calculations}
We focus here on the interaction between a symmetric top molecule, CH$_3$NC or its isomer CH$_3$CN, with a structureless helium atom ($^1$S) in their respective electronic ground states.
In the present work, we consider these molecules to be rigid rotors, leading to a three-dimensional potential energy surface (3D-PES). Since both CH$_3$CN and CH$_3$NC possess a low energy vibrational bending
mode (365 and 270 cm$^{-1}$, respectively), we note that the PES should eventually include a dependence on this vibrational coordinate to treat collisional excitation except at low temperatures.

\begin{figure}
\centering
	\includegraphics[width=0.6\columnwidth]{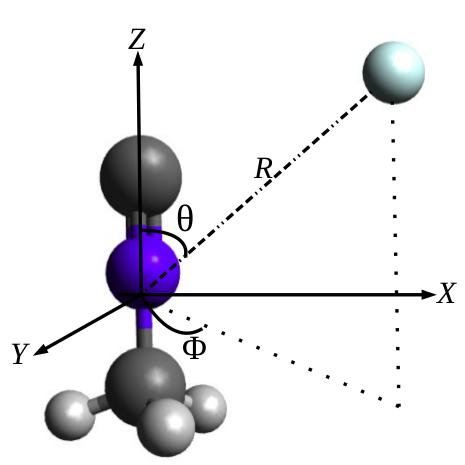}
	\caption{Jacobi coordinates used to describe the interaction of CH$_{3}$NC-He van der Waals complex. The origin of the reference
	frame is at the CH$_3$NC center-of-mass.
	The angles are defined so that for $\theta=0^{\circ}$ He approaches along the NC/CN end of the molecule and for $\phi=0^{\circ}$ the He atom lies in the plane defined by (NC/CN)CH atoms on the side of the H atom. }
	\label{CH3CN-H2-CM}
\end{figure}

The coordinate system used in this work is presented in Figure~\ref{CH3CN-H2-CM}. 
The origins of the reference frames are the centers of mass of CH$_3$NC and CH$_3$CN respectively and the $C_3$ principal axis of CH$_3$ is located along the $z$ axis. Consequently, three degrees of freedom are needed to describe the interaction
potential of CH$_3$NC-He and CH$_3$CN-He.
We use the following coordinates: $(i)$ the distance $R$ between the center of mass of the target systems to the helium atom, $(ii)$ the polar angle $\theta$, which describes the position of the helium atom with respect
to the $z$-axis, and $(iii)$ the azimuthal angle $\phi$ that describes the rotation of the He atoms around the $z$ axis.

The 3D-PESs are calculated within the rigid rotor approximation and the internal coordinates of methyl (iso)cyanide are described using the experimental ground vibrational geometries as follows:
$r$(N$\equiv$C)=1.167\AA, $r$(C--N)=1.416\AA, $r$(C--H)=1.093\AA, and $\angle$(HCH)= 109.5$^{\circ}$ for CH$_3$NC \cite{vijay1996ab} and $r$(C$\equiv$N)=1.156\AA, $r$(C--C)=1.457\AA,
$r$(C--H)=1.087\AA, and $\angle$(HCC)= 110$^{\circ}$ for CH$_3$CN \cite{le1992rotational}. Consequently, the PESs were computed
using 52 values of $R$ ranging from 3.75 to 50a$_0$, 19 values of $\theta$ uniformly spanning the range [0$^{\circ}$-180$^{\circ}$] by steps of 10$^{\circ}$ and 7 values of $\phi$ angles in the range
[0-60$^\circ$]. These parameters lead to a set of 6916 energy points.

The \textit{ab initio} energies were computed using the {\small MOLPRO} package \cite{werner2015molpro} and treated in the C$_1$ symmetry group. We use the explicitly correlated coupled cluster method with inclusion of single,
double and non-iterative triple excitations [CCSD(T)-F12a] \cite{knizia2009simplified} in conjunction with the augmented
correlation-consistent polarized valence triple zeta (aug-cc-pVTZ) basis sets of Dunning \cite{dunning1989gaussian}.
The CCSD(T)-F12a method with the aug-cc-pVTZ basis set should lead to an accurate PES. Indeed, it has been shown that CCSD(T)-F12 with AV$n$Z basis sets yields an accuracy comparable to standard CCSD(T) with AV($n$+1)Z basis sets with reduced computational time \cite{knizia2009simplified}.

We take into account the correction for the basis set superposition errors (BSSE) using the counterpoise procedure of \citet{boys1970calculation} according to the following expression:
\begin{equation}\label{eq:1}
 V(R,\theta,\phi)=V_{\rm{Mol-He}}(R,\theta,\phi)-V_{\rm{Mol}}(R,\theta,\phi)-V_{\rm{He}}(R,\theta,\phi)
\end{equation}
where $V_{\rm{Mol-He}}$ is the global electronic energy of CH$_3$NC-He and CH$_3$CN-He systems, and the last two terms are the energies of the two
fragments, all performed using the full basis set of the total system.

A disadvantage of the CCSDT(T)-F12a method is that it is not size consistent, therefore,
the PESs computed with this approach were shifted by subtracting the asymptotic value of the interaction potential, which is equal to $-$4.09 cm$^{-1}$ for CH$_3$NC-He and $-$4.80 cm$^{-1}$ for CH$_3$CN-He.

\subsection{Analytical fit}\label{sec:2}

As required by {\small MOLSCAT} \cite{molscat95} and {\small HIBRIDON} \cite{Hibridon} codes that were used for the scattering calculations, we fitted our \textit{ab initio} data to an analytical expansion 
in order to produce the global PES $V(R,\theta,\phi)$  according to the following functional form \cite{green1976rotational}:
\begin{equation}
V(R,\theta,\phi)=\sum_{l=0}^{l_{max}}\sum_{m=0}^{l}V_{lm}(R)Y_{l}^{m}(\theta,\phi)\label{eq:expansion}
\end{equation}
Taking into consideration the property of spherical harmonics, the 3D-PESs can be written as: 
\begin{equation}
V(R,\theta,\phi)=\sum_{l=0}^{l_{max}}\sum_{m=0}^{l}V_{lm}(R)\frac{Y_{l}^{m}(\theta,\phi)+(-1)^{m}Y_{l}^{-m}(\theta,\phi)}{1+\delta_{m,0}}
\end{equation}
where $V_{lm}(R)$ and $Y_{l}^{m}(\theta,\phi)$ denote the radial coefficients to be computed and the normalized spherical harmonics respectively, and
$\delta_{m,0}$ is the Kronecker symbol.

\begin{figure}
\centering
	\includegraphics[width=1\columnwidth]{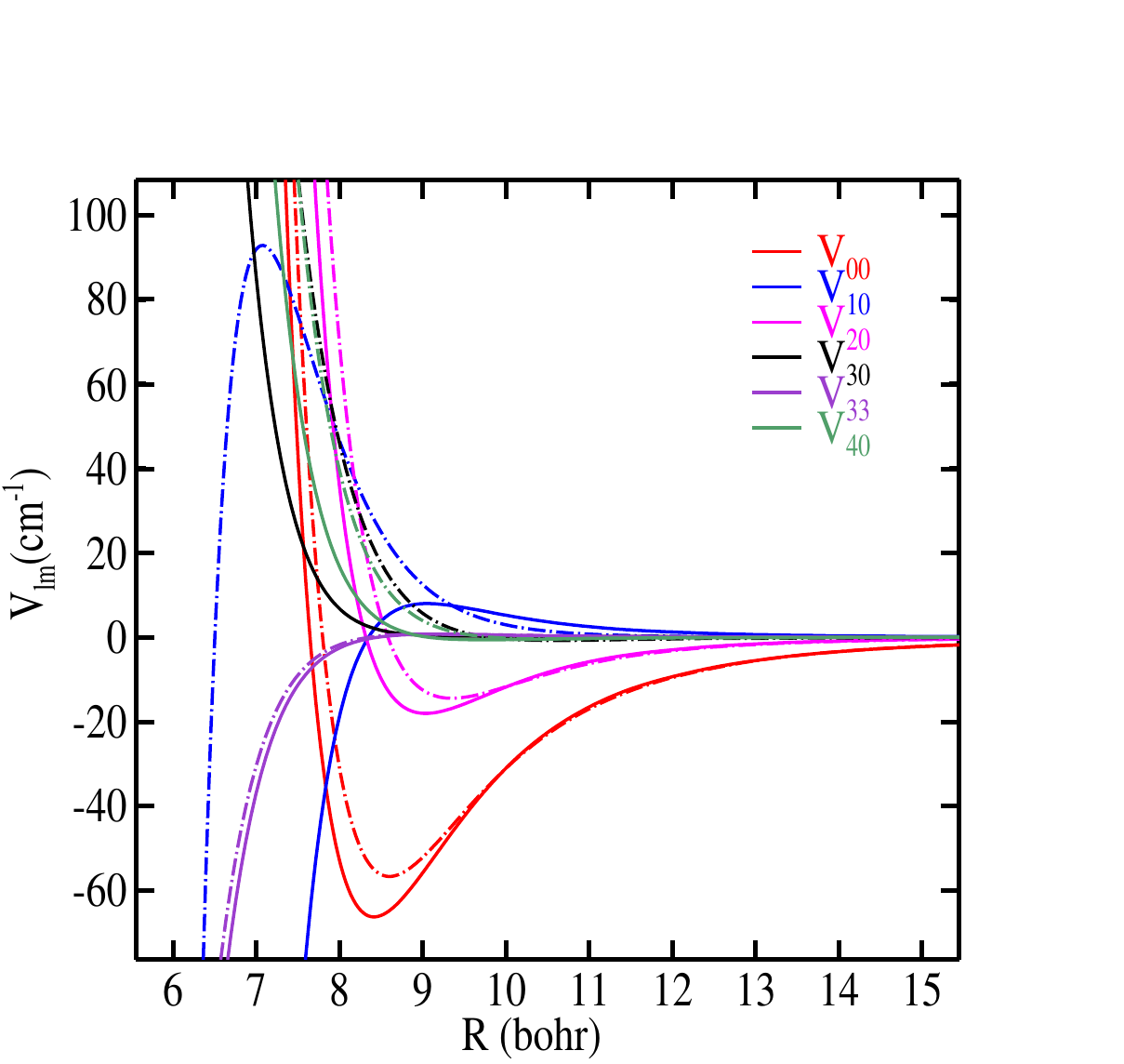}
	\caption{Dependence on $R$ of the first $V_{lm}(R)$ components for CH$_3$CN-He (solid lines) and CH$_3$NC-He (dashed lines) with 0 $\leq l \leq$ 4.}\label{coeff}
\end{figure}

Due to the C$_{3v}$ symmetry of the CH$_3$NC molecule and its isomer, the parity of the expansions is constrained to $m$ multiple of 3 ($m=3n$ , $n$ integer). Hence, as we used 19 values
of $\theta$ and 7 values of $\phi$ angles to compute the \textit{ab initio} PESs, we can compute radial coefficients up to ($l_{max}$,$m_{max}$)=$(18,18)$, leading to a set of 70 angular functions.
Scattering calculations require continuous radial coefficients, and for each value of $R$ we used a least-square procedure to generate the $V_{lm}(R)$ radial coefficients. 
This development reproduces the global \textit{ab initio} potential with a root means square deviation (rms) of 1.0 cm$^{-1}$. The rms is much larger in the short range due to the larger anisotropy of the PES, e.g., at $R = 4.75a_0$ it is equal to 224 cm$^{-1}$. The rms becomes less than 1 cm$^{-1}$ for distances $R \geq 6a_0$ and decreases with increasing $R$, which shows that the minimum of the PES and the long-range region are adequately described.
For $R$ $\geq$ 30 bohr, we extrapolated the long-range potential using an inverse exponent expansion implemented in the {\small MOLSCAT} code.
We illustrate in figure~\ref{coeff} the dependence on $R$ of the radial coefficients up to $l$= 4 for CH$_3$NC and CH$_3$CN. 
Here, $V_{00}$ represents the isotropic potential, responsible for elastic collisions, while terms with $l \geq$ 1 describe the anisotropic part of the PES responsible for inelastic collisions.
As a first outcome of the fitting, we note that the isotropic term $V_{00}$ presents a deeper well for CH$_3$CN-He (65.8 cm$^{-1}$) than CH$_3$NC-He (56.2 cm$^{-1}$).
In addition, we note that for a given distance, the $V_{lm}$ with even values of $l$ are larger for CH$_3$CN than those for CH$_3$NC and $V_{lm}$ with odd value of $l$ are deeper for CH$_3$NC compared to those for CH$_3$CN.
This is expected to have an effect on the propensity rules in the rotational excitation, as shall be further discussed below.

\subsection{Analysis of the PES}
For CH$_3$NC-He, a global minimum with $D_e= 58.61$ cm$^{-1}$ exists at $\phi$=60$^{\circ}$, $\theta$ = 100$^{\circ}$ and $R=6.05$ bohr while a local minimum of 46.77 cm$^{-1}$ is located at $\theta$ = 180$^{\circ}$ and $R=8.5$ bohr.  

For CH$_3$CN-He, the PES present a global minimum with a depth of $D_e$= 55.10 cm$^{-1}$
for the configuration $\phi$=60$^{\circ}$,
$\theta$ = 100$^{\circ}$ and $R=6.15$ bohr while the local minimum of 38.47 cm$^{-1}$ is located at $R=8.6$ bohr and $\theta$=180$^{\circ}$. We note that the PESs for these isomers share the same qualitative behavior previously observed for molecules with a threefold symmetry axis interacting with rare gas atoms  \cite{Gubbels2012, Loreau2014b, loreau2015scattering} where the global minimum occurs at $\phi$=60$^{\circ}$, \textit{i.e.} with the helium atom located between two hydrogen atoms. However, the global minimum as well as the local minimum of CH$_3$NC-He are deeper than those of  CH$_3$CN-He. 

The global and local minima are separated by a transition state of $-$29.05 cm$^{-1}$ located at $\theta$=139.5$^{\circ}$ and $R=8.25$ bohr for CH$_3$NC-He, while for CH$_3$CN-He this transition state is at $R=8.5$ bohr and $\theta$=143$^{\circ}$ with an energy of -28.51 cm$^{-1}$.

\begin{figure*}
\centering
{\label{a}\includegraphics[width=.49\linewidth]{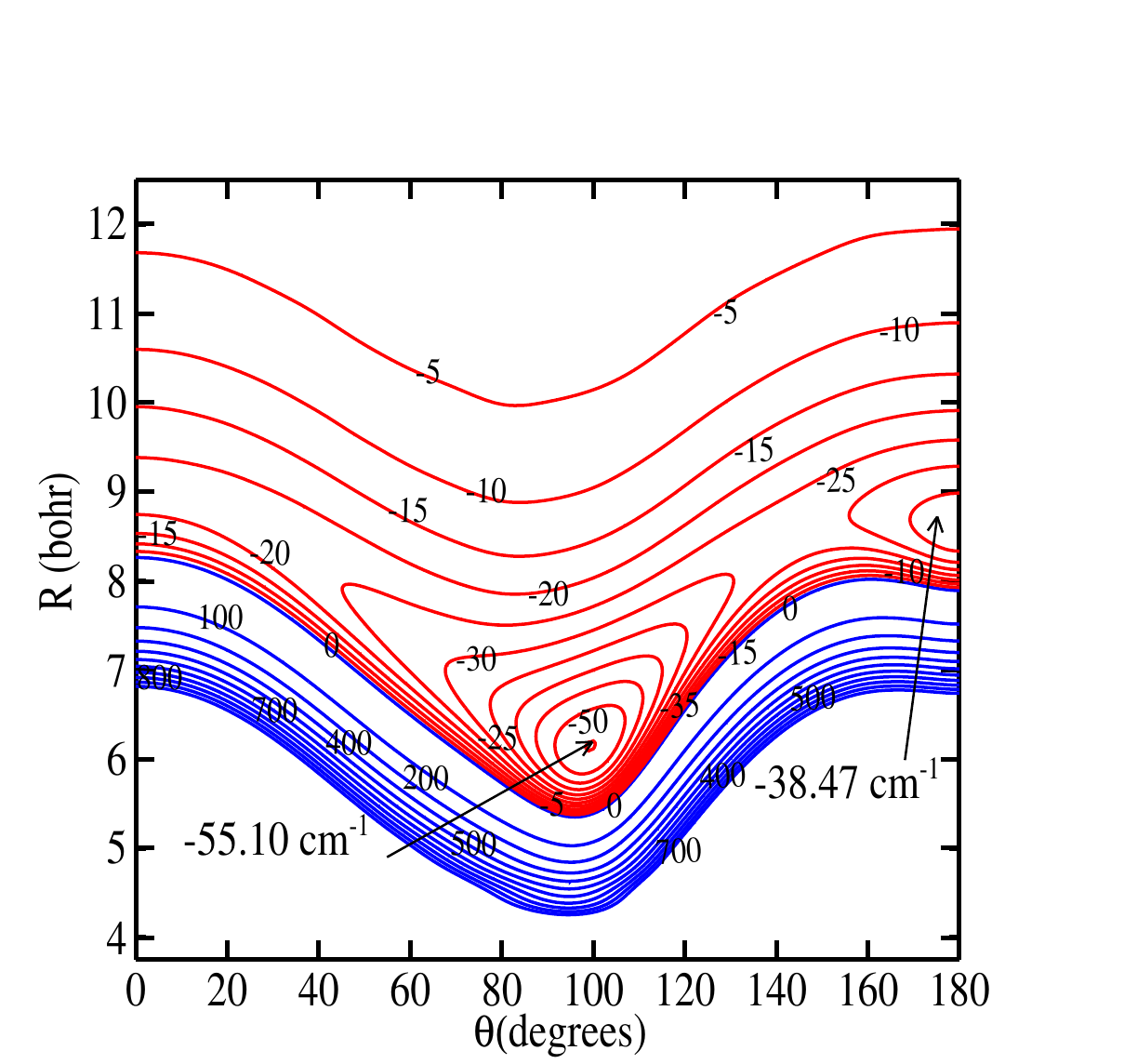}}
{\label{b}\includegraphics[width=.49\linewidth]{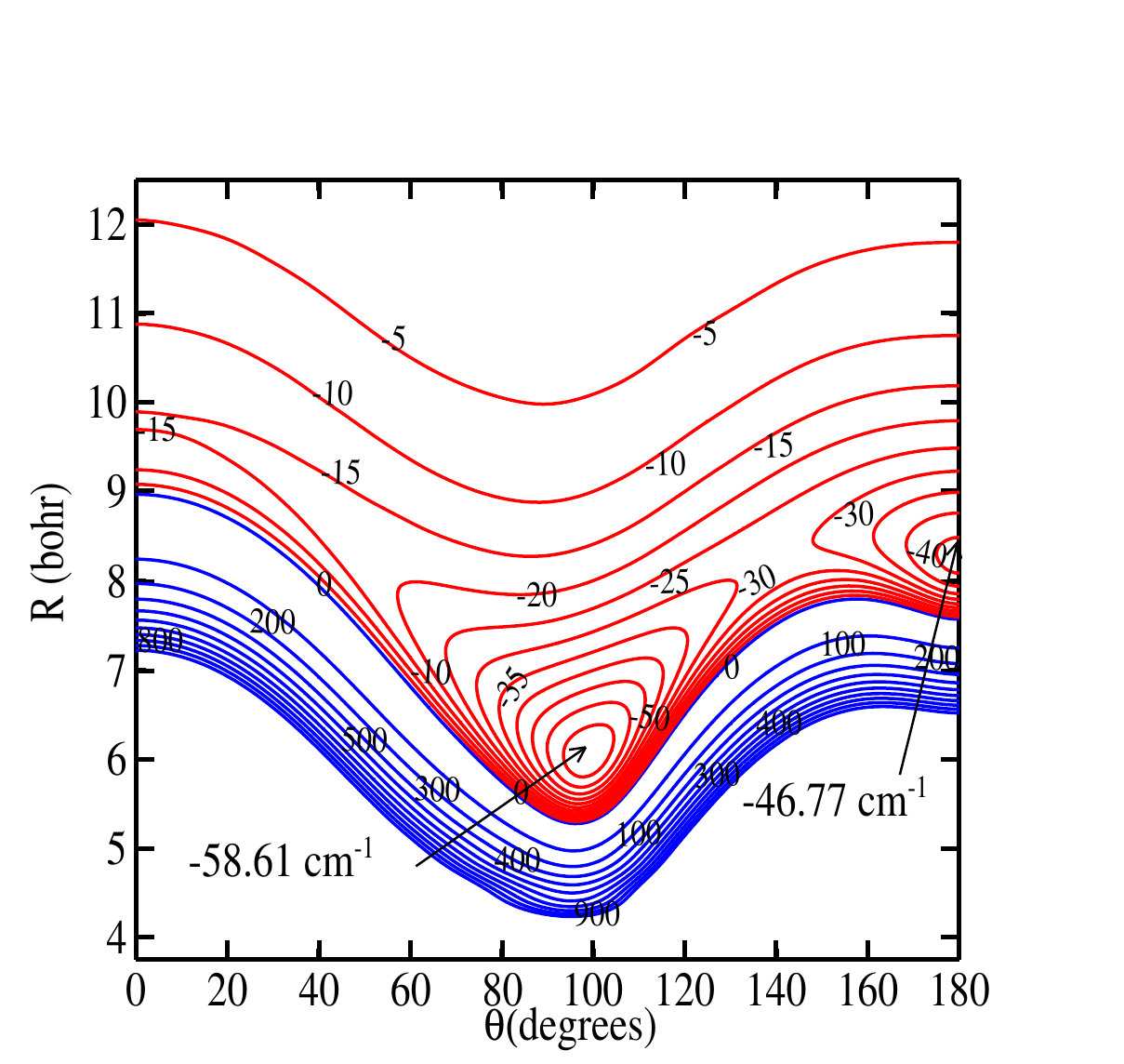}}
{\label{c}\includegraphics[width=.49\linewidth]{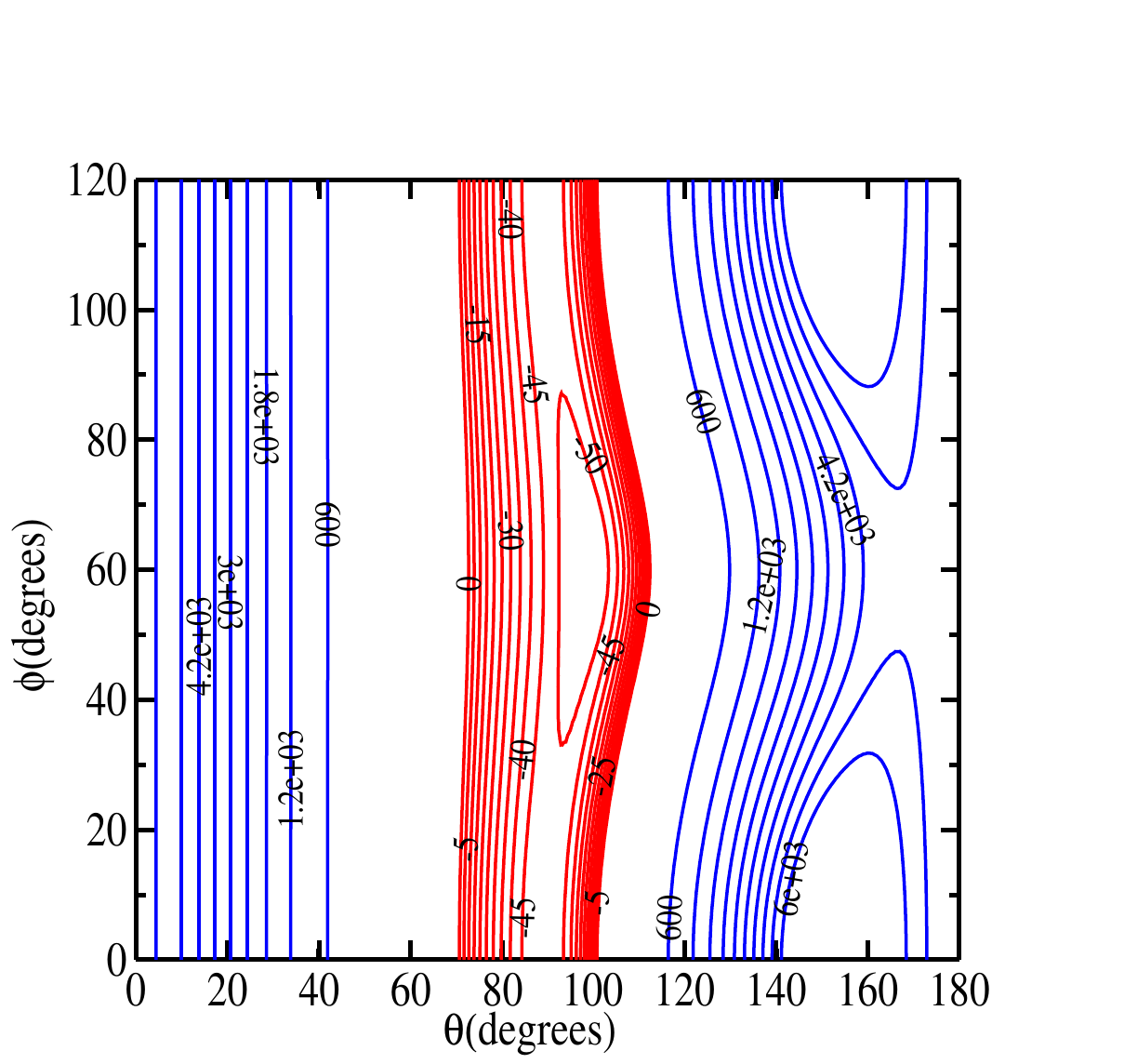}}
{\label{c}\includegraphics[width=.49\linewidth]{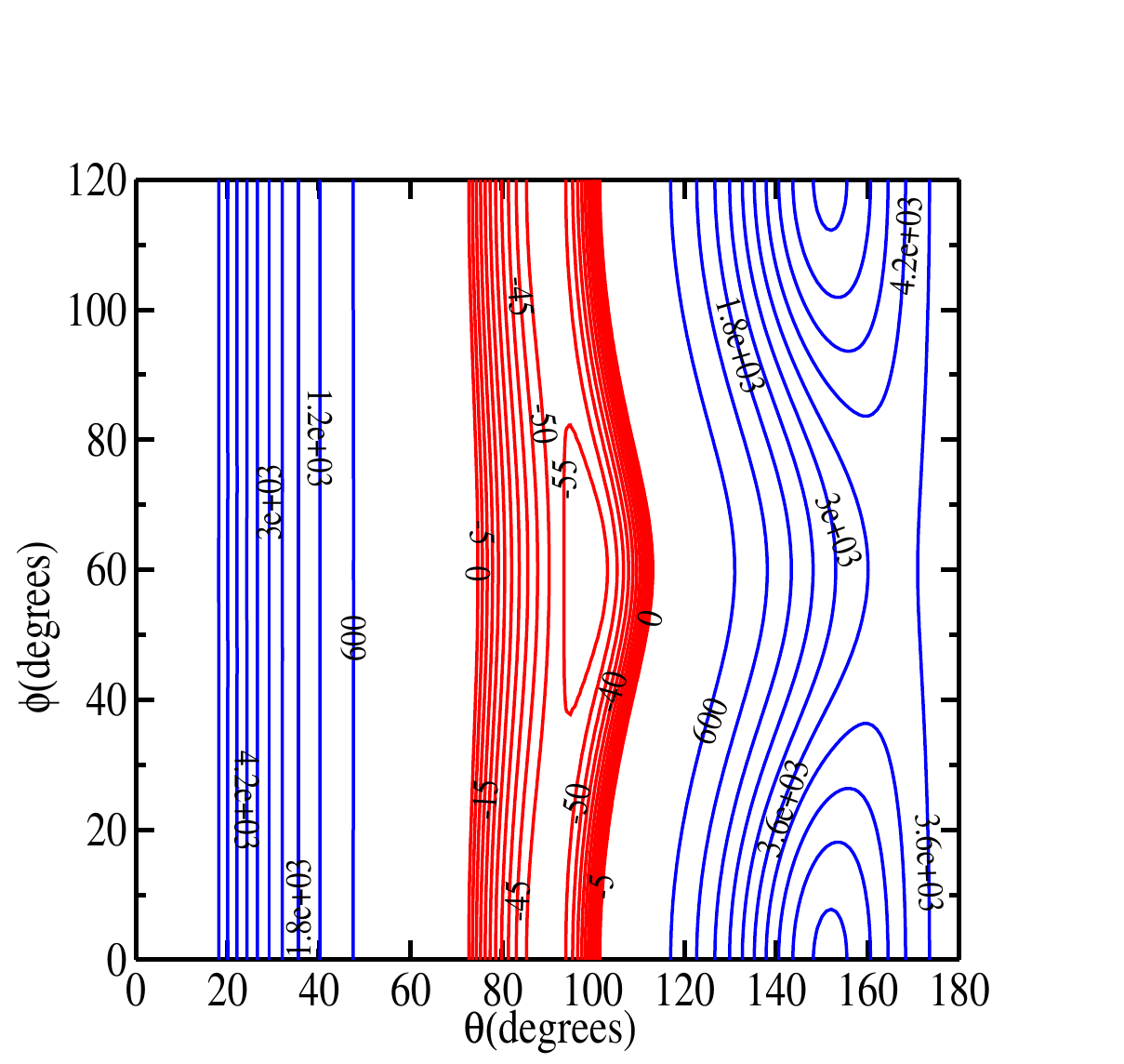}}
\caption{Two-dimensional contour plots of the interaction potential of the CH$_3$CN-He (left panels) and CH$_3$NC-He (right panels) van der Waals complex. Top panels
depict the 3D-PESs as a function of $\theta$ and $R$ at $\phi = 60^{\circ}$, while the bottom panels show the PESs as a function of $\phi$ and $\theta$ at $R=6.1$ a$_0$. For each panel, the blue (red) contours represent the positive (negative) parts of the potential (in unit of 
cm$^{-1}$). }\label{3D-PES}
\label{fig-PES} 
\end{figure*}

We illustrate in Fig~\ref{3D-PES} two-dimensional cuts of the interaction potentials as a function of two Jacobi coordinates, while the third one is held fixed at its equilibrium
values in the CH$_3$NC-He and CH$_3$CN-He minimum.
The variation in those cuts shows a stronger anisotropy of the interaction potential of CH$_3$NC-He compared to CH$_3$CN-He along the $\theta$ coordinate.

We compared the new 3D-PES of CH$_3$CN-He with the earlier CH$_3$CN-He PES computed using the automated interpolating moving least squares methodology \cite{khalifa2020rotational}. We conclude
that the two PESs are qualitatively and quantitatively similar, the energies and locations of the stationary states being almost identical. As an example, the global minimum was found at $\phi$=60$^{\circ}$, $\theta$=100.4$^{\circ}$ and $R$=6.14 bohr with an energy of 55.16  cm$^{-1}$, to be compared to our values of 55.10 cm$^{-1}$ at $\phi$=60$^{\circ}$, $\theta=100.0^{\circ}$ and $R$=6.15 bohr.
%%%% JL: give values here

For a better appreciation of the topography of the PESs, we present also 2D-cut of the 3D-PESs as a function of $\theta$ and $\phi$ for $R=6.1$ bohr in the bottom panels of Fig. \ref{3D-PES}. This type of plot offers a unique overview as it includes all minima and the barriers between them.
Its should also be noted that the potential for CH$_3$NC-He is much more repulsive than that of CH$_3$NC-He at small and large values of $\theta$, while the two PESs are similar in the region of the minimum. This leads to a higher anisotropy for the CH$_3$NC-He potential, as already alluded to above.

\section{Bound states}\label{sec:bound}

We have computed the energies of bound vibrational levels of the CH$_3$CN--He and CH$_3$NC--He van der Waals complexes in order to determine the
dissociation energy $D_0$ of the complexes. A set of basis functions formed from the product of stretching and body-fxed angular functions was employed \cite{green1976rotational}.
The $R$-space was spanned by a set of 16 distributed Gaussians \cite{Hamilton1986a}. The width of the Gaussians is 1.33
a$_{0}$ and the range is spanned from 5 to 11 $a_{0}$. The HIBRIDON code \cite{Hibridon} was employed to determine the
energies of the bound vibrational levels.

Due to the permutation of the three hydrogen atoms, the rotational levels of CH$_3$NC and CH$_3$CN are divided into two groups: $A-$CH$_3$NC/CN ($ortho$) and $E$-CH$_3$NC/CN ($para$) corresponding
to the $A_1$ or $A_2$, and $E$ irreducible representations of the C$_{3\text{v}}$ point group. In fact, each hydrogen nucleus possesses a nuclear spin 1/2, hence, 
if we couple the three hydrogen nuclear spins, we obtain a total nuclear spin equal to $I$= 3/2 ($ortho-$CH$_3$NC/CN) or $I$= 1/2 ($para-$CH$_3$NC/CN).
The rotational states of $para-$CH$_3$NC/CN correspond to $k= 3n\pm 1$, while those of $ortho-$CH$_3$NC/CN correspond to $k= 3n$ for  ($n$ being integer).

Separate calculations of the dissociation energies of van der Waals complexes of the $ortho$ and $para$ levels of CH$_3$CN--He and CH$_3$NC--He were carried
out. The dissociation energy $D_0$ of $ortho$-CH$_3$CN--He and $ortho$-CH$_3$NC--He complexes were found to almost equal:  $D_0(ortho$-CH$_3$CN--He) = 18.64 cm$^{-1}$ and 
$D_0(ortho$-CH$_3$NC--He) = 18.65 cm$^{-1}$. Levels up to $j=24$ were required in the rotational basis set for convergence of the dissociation energy.
No bound levels were found for the $para$-CH$_3$CN--He and $para$-CH$_3$NC--He complexes. The dissociation energies are summarized in Table \ref{table_D0}.

\begin{table}
\centering
\caption{Dissociation energies $D_0$ (in cm$^{-1}$) of the $ortho$ and $para$ levels of CH$_3$CN--He and CH$_3$NC--He}\label{table_D0}
\begin{tabular}{ccc}
\hline
\hline
 & CH$_3$CN-He & CH$_3$NC-He    \\
\hline
$ortho$ & 18.64 & 18.65   \\
$para$ & 0 & 0  \\
\hline
\hline
\end{tabular}
\end{table}

\section{Scattering calculations}\label{sec:scat}

\subsection{Spectroscopy of CH$_3$CN and CH$_3$NC}

CH$_3$CN and CH$_3$NC are prolate symmetric top molecules.
The relevant Hamiltonian of such a symmetric rotor is given by:
\begin{equation}
 H_{rot}=\frac{\hbar^2}{2I_b}j^2+\hbar^2(\frac{1}{2I_a}-\frac{1}{2I_b})j_a^2
\end{equation}
where $j$ is a rotational quantum number that satisfies the relation $j^2$=$j^2_a+j^2_b+j^2_c$ and $I_a$ and $I_b$ are the principal moments of inertia.
The rotational wave functions $\vert jkm \rangle $ of both isomers are defined by three quantum numbers, where $k$ denotes 
the projection of $j$ along the $a$-axis of the body-fixed reference and $m$ is its projection on the $z$-axis of the space-fixed frame of reference.
The  eigenvalues is the energies of the rotational levels, given by :
\begin{equation}
E_{j,k}=Bj(j+1)+(A-B)k^2
\end{equation}
where $B$ and $A$ are the rotational constants which are equal to 0.3353 cm$^{-1}$ and 5.2420 cm$^{-1}$ for CH$_3$NC \cite{margules2001ab} and 0.3068 cm$^{-1}$ and 5.2470 cm$^{-1}$
for CH$_3$CN \cite{remijan2007alma} respectively.

\begin{figure}
\centering
	\includegraphics[width=0.9\columnwidth]{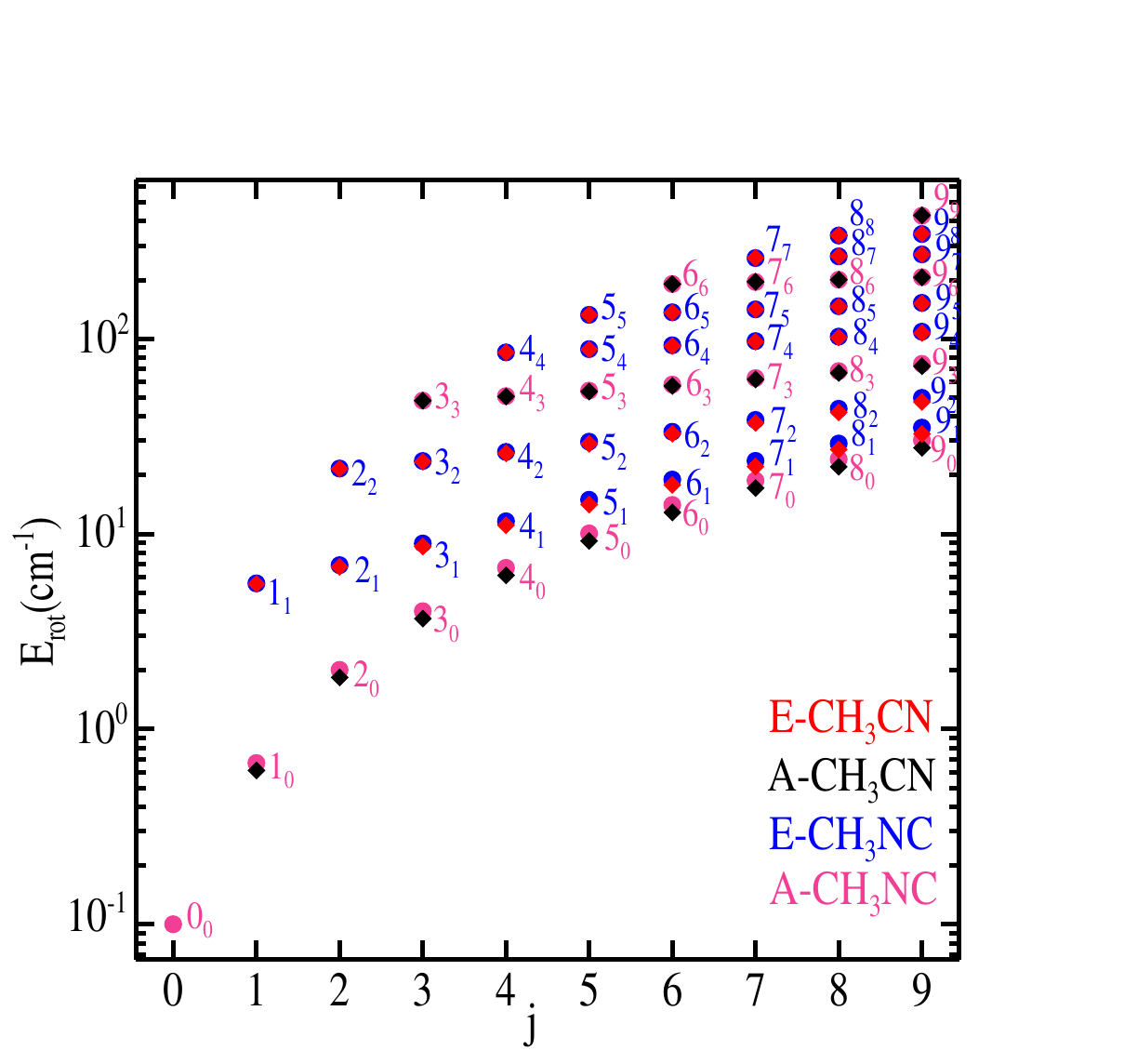}
	\caption{Rotational levels of $para-$ and $ortho-$CH$_3$NC (circles) and CH$_3$CN (diamonds) up to $j_k = 9_9$.}\label{Levels}
\end{figure}

For illustration, we present in figure~\ref{Levels} the rotational energy diagram associated to the CH$_3$NC and CH$_3$CN levels up to $j=9$.
It must be emphasized that the $ortho-$ and $para-$ states cannot be interconverted either by radiative transitions, nor by inelastic collisions: hence, the conversion from $ortho-$CH$_3$NC/CN to
$para-$CH$_3$NC/CN is forbidden (and vice versa) and the scattering calculations can be performed separately for each nuclear spin species.

The small value of the rotational constant $B$ leads to a high density of rotational levels even at low energy. In astrophysical environments, CH$_3$CN and CH$_3$NC are indeed often observed through transitions involving highly excited states, e.g. 17$_k$ to 16$_k$, $k$= 0,...,7 and 18$_k$ to 17$_k$, $k$= 0,...,7) for CH$_3$NC \cite{calcutt2018alma}. This implies that the theoretical study of rotational excitation of methyl (iso)cyanide requires a large basis set of rotational levels, which in turn limit the possibility of using fully quantum-mechanical dynamics methods to low collision energies.

\subsection{Cross section computations}

In this work, we present cross sections for rotationally inelastic transitions up to $E_{tot} \leq$ 100 cm$^{-1}$, computed using the time-independent close-coupling (CC) method.
In order to perform the scattering calculations of CH$_3$NC-He and CH$_3$CN-He colliding systems, we introduced radial coefficients $V_{lm}(R)$ in MOLSCAT code using the VSTAR routine.
The diabatic log-derivative propagator is used in order to solve the coupled equations \cite{manolopoulos1986improved}. Some of the calculations were carried out with the HIBRIDON suite of programs \cite{Hibridon}.  We do not take into consideration the hyperfine
excitation in both isomers because the component splitting is not yet resolved by the astrophysical observations.

We focus here on the low energy ($E \leq 100$ cm$^{-1}$) scattering of CH$_3$NC-He and CH$_3$CN-He. 
In order to properly converge collisional cross sections, preliminary tests were performed in order to find the optimal parameters at several energies. This includes the integration boundaries of the propagator, R$_{min}$ and R$_{max}$, which were fixed at 2.5 and 50 bohr respectively; the number of integration steps, taken as 100 for $E_{tot} \leq$ 50 cm$^{-1}$ and 70 for $E_{tot} \in$ [50,100] cm$^{-1}$;
as well as the size of the rotational basis set which includes all open channels and some closed channels. Rather than including all rotational states up to a given $j_{\max}$, it is more convenient to use a cutoff based on energy and include all states up to that value.
For $ortho-$CH$_3$NC, it was found necessary to include the 35 lowest levels (up to $j_k=20_0$ with an energy of 140.8 cm$^{-1}$) for total energies  $E_{tot} \leq$ 50 cm$^{-1}$ and 55 levels (up to $j_k=26_0$ with an energy of 235.4 cm$^{-1}$)  for 50 $\leq E_{tot} \leq$ 100 cm$^{-1}$, while for $ortho-$CH$_3$CN we need the first 55 (up to  $j_k=12_6$, energy of 225.6 cm$^{-1}$) and 65 levels (up to $j_k=29_3$, energy of 311.4 cm$^{-1}$), respectively.
For $para-$CH$_3$NC-He, the first 65 rotational levels (up to  $j_k=13_5$, with an energy of 183.7 cm$^{-1}$) are necessary to converge cross sections for $E_{tot} \leq$ 50 cm$^{-1}$, and 75 states (up to  $j_k=20_4$ with an energy of 219.3 cm$^{-1}$) for $E_{tot} \in$ [50,100] cm$^{-1}$, while for $para-$CH$_3$CN, we take the first 78 rotational levels (up to $j=18_7$, with an energy 346.9 cm$^{-1}$) for $E_{tot} \le100$ cm$^{-1}$.

The energy grid was chosen as dense enough to describe resonances at the various thresholds corresponding to the opening of rotational levels, and the spacing is significantly increased outside the resonance zones. For $0< E_{tot} \leq 50$ cm$^{-1}$ we used a step size of 0.2 cm$^{-1}$, while for $50< E_{tot} \leq 100$ cm$^{-1}$ the step size was 0.5 cm$^{-1}$.
The reduced mass is taken at $\mu$= 3.646815237 atomic mass unit (isotopes $^{12}$C,$^{14}$N, $^{1}$H and $^{4}$He).

\begin{figure}[!ht]
\centering
%    \textbf{CH$_3$NC-He \hspace{5cm} CH$_3$CN-He}
    \begin{minipage}{.5\textwidth}
\includegraphics[width=0.8\textwidth]{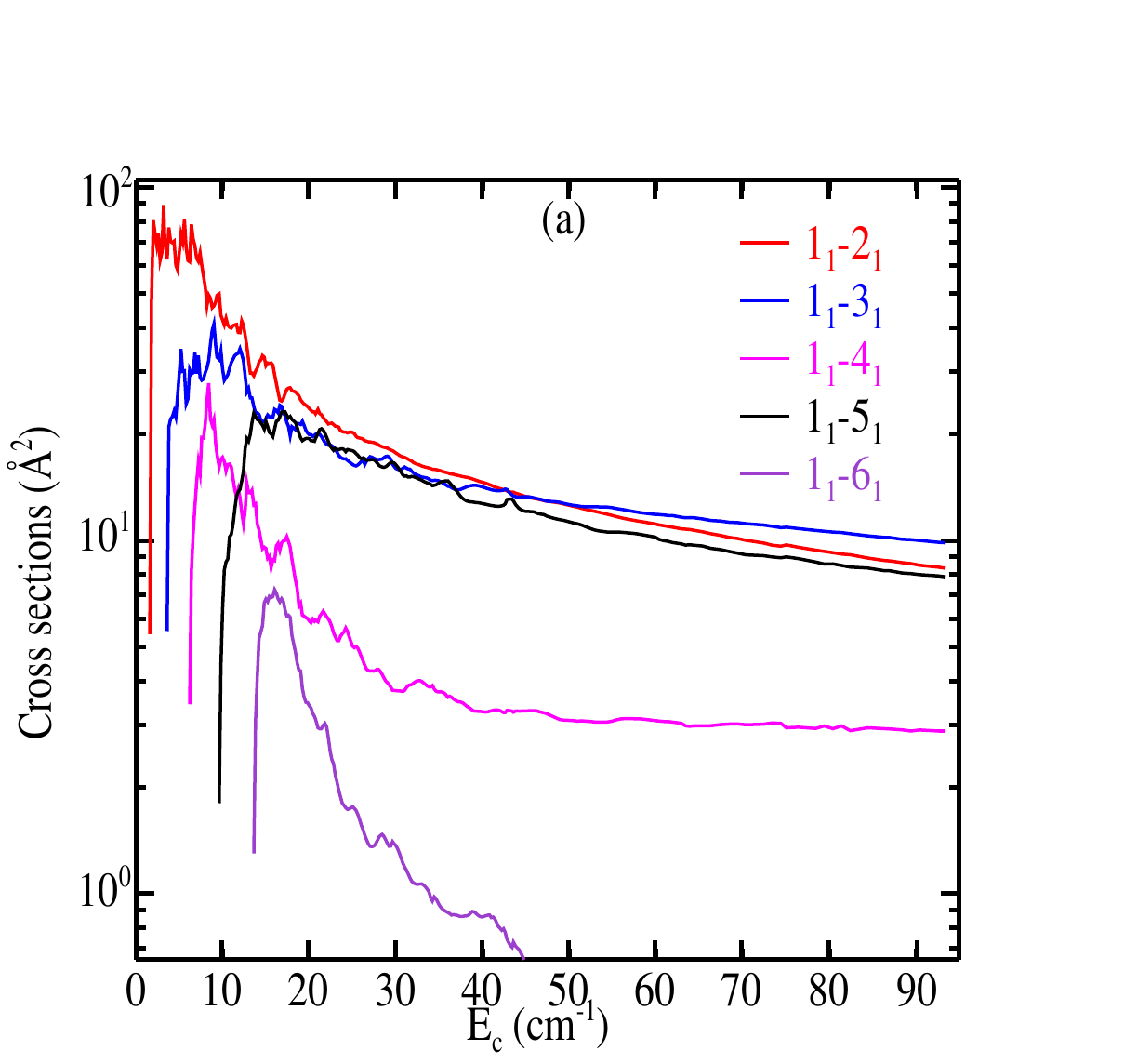}
\includegraphics[width=0.8\textwidth]{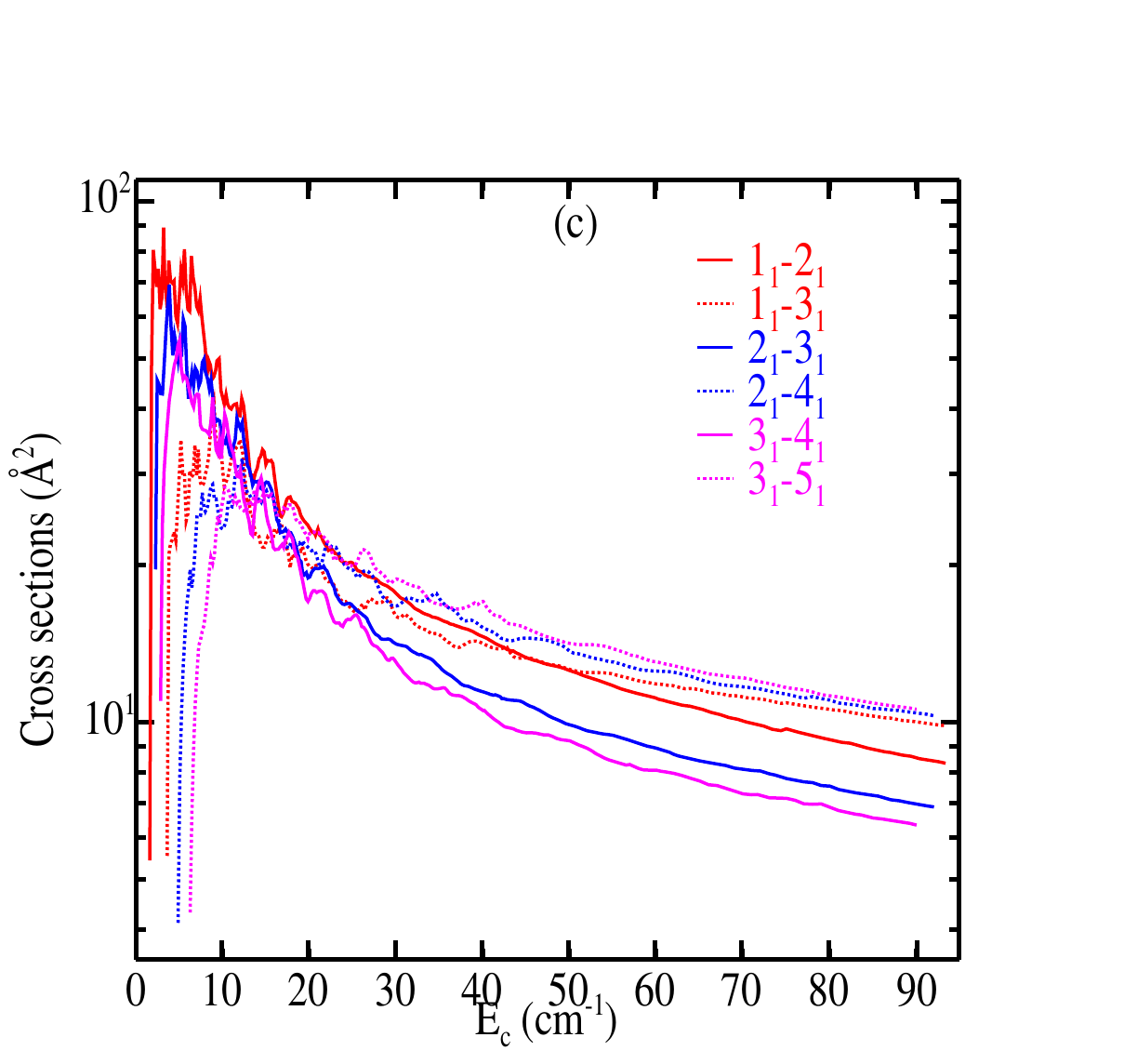}
\includegraphics[width=0.8\textwidth]{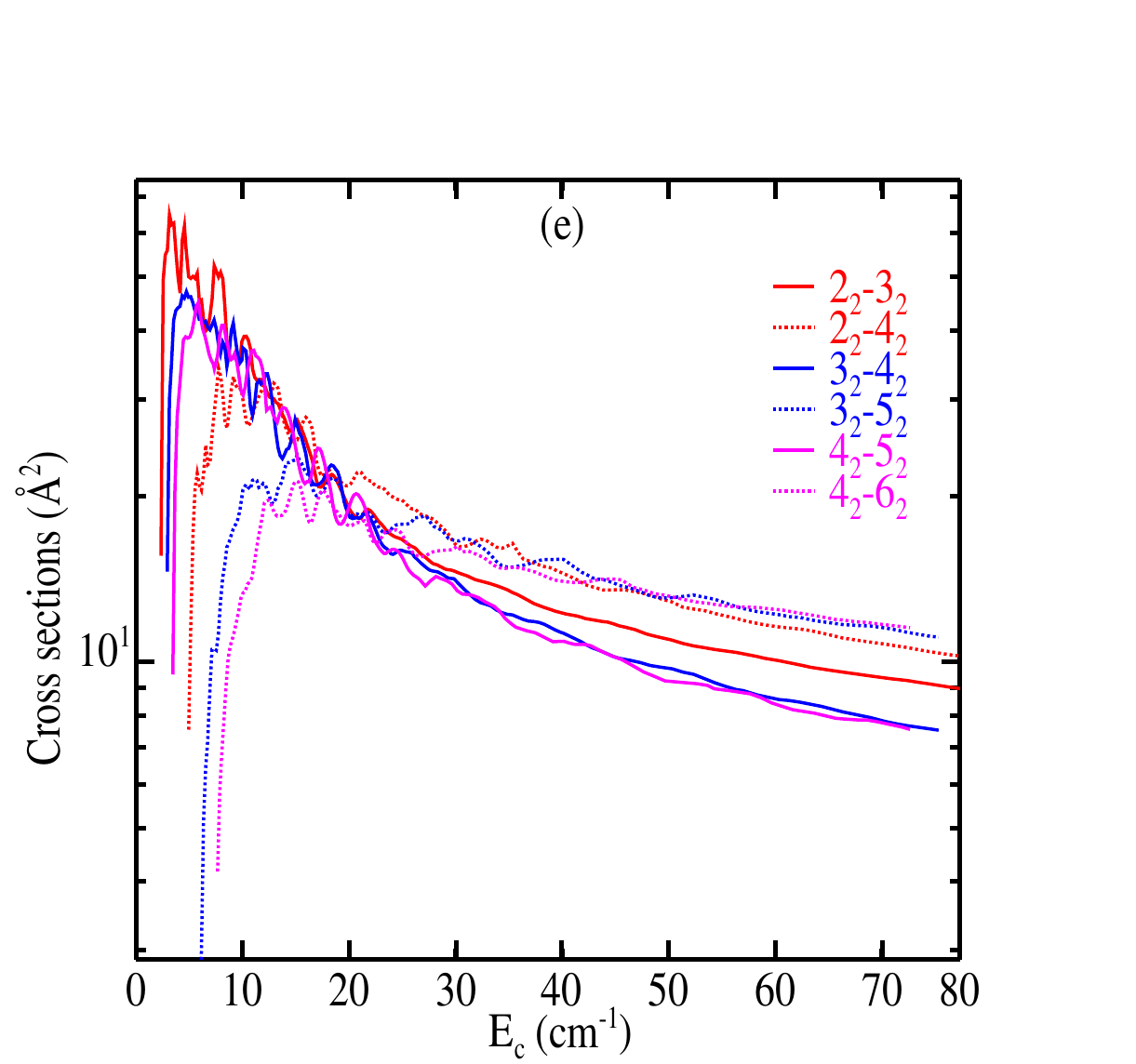}
\end{minipage}%
\begin{minipage}{.5\textwidth}
\includegraphics[width=0.8\textwidth]{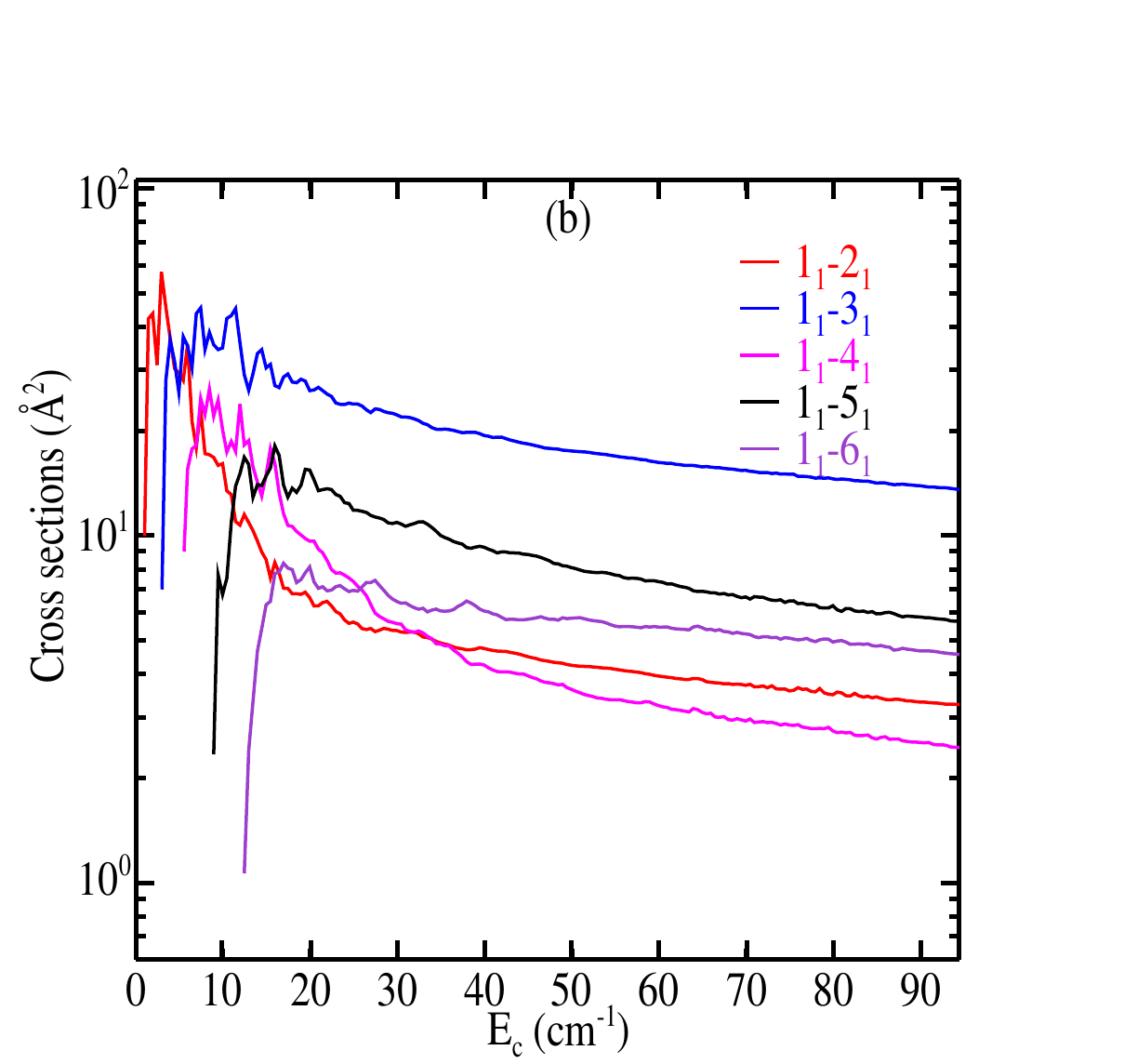}
\includegraphics[width=0.8\textwidth]{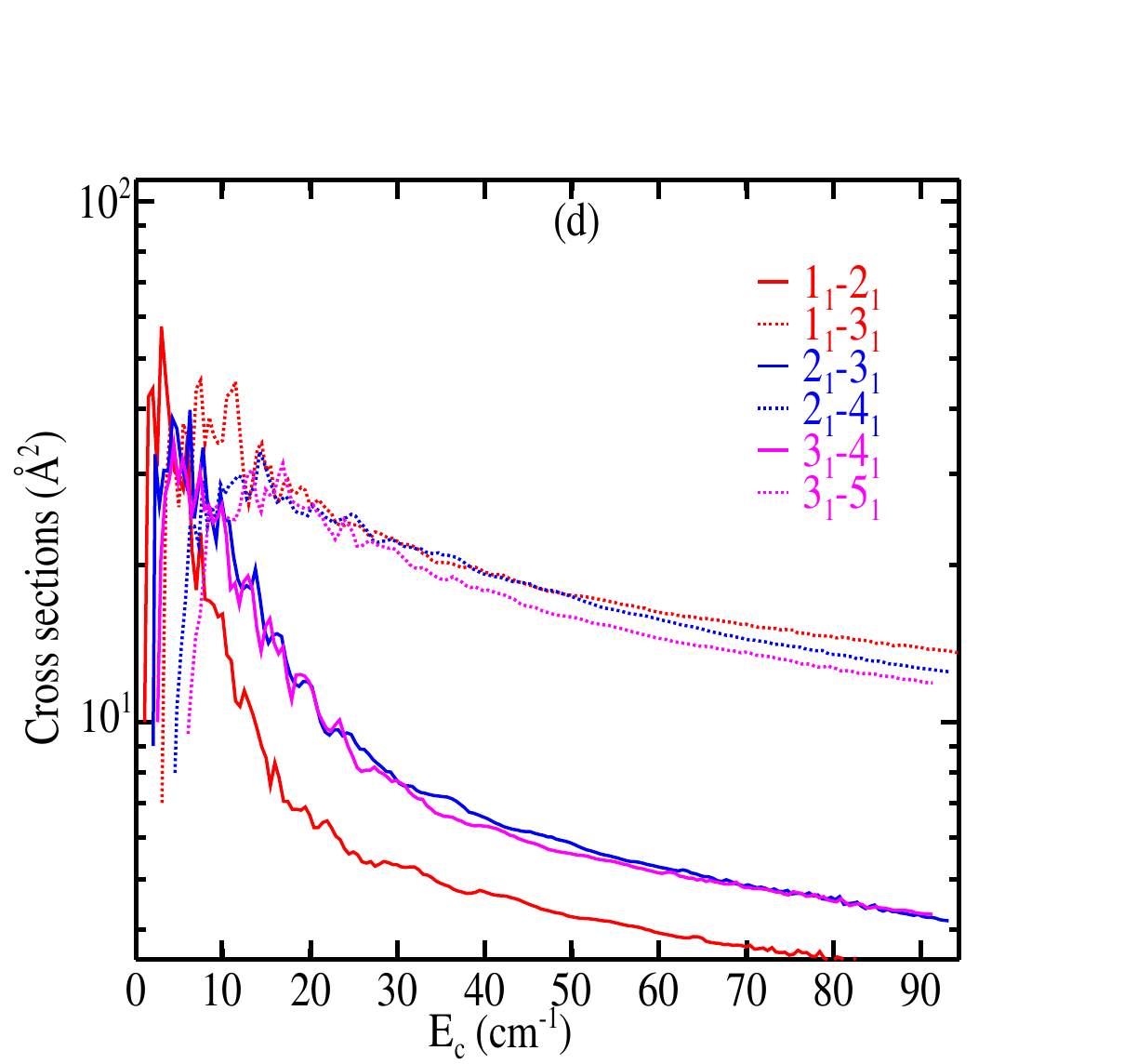}
\includegraphics[width=0.8\textwidth]{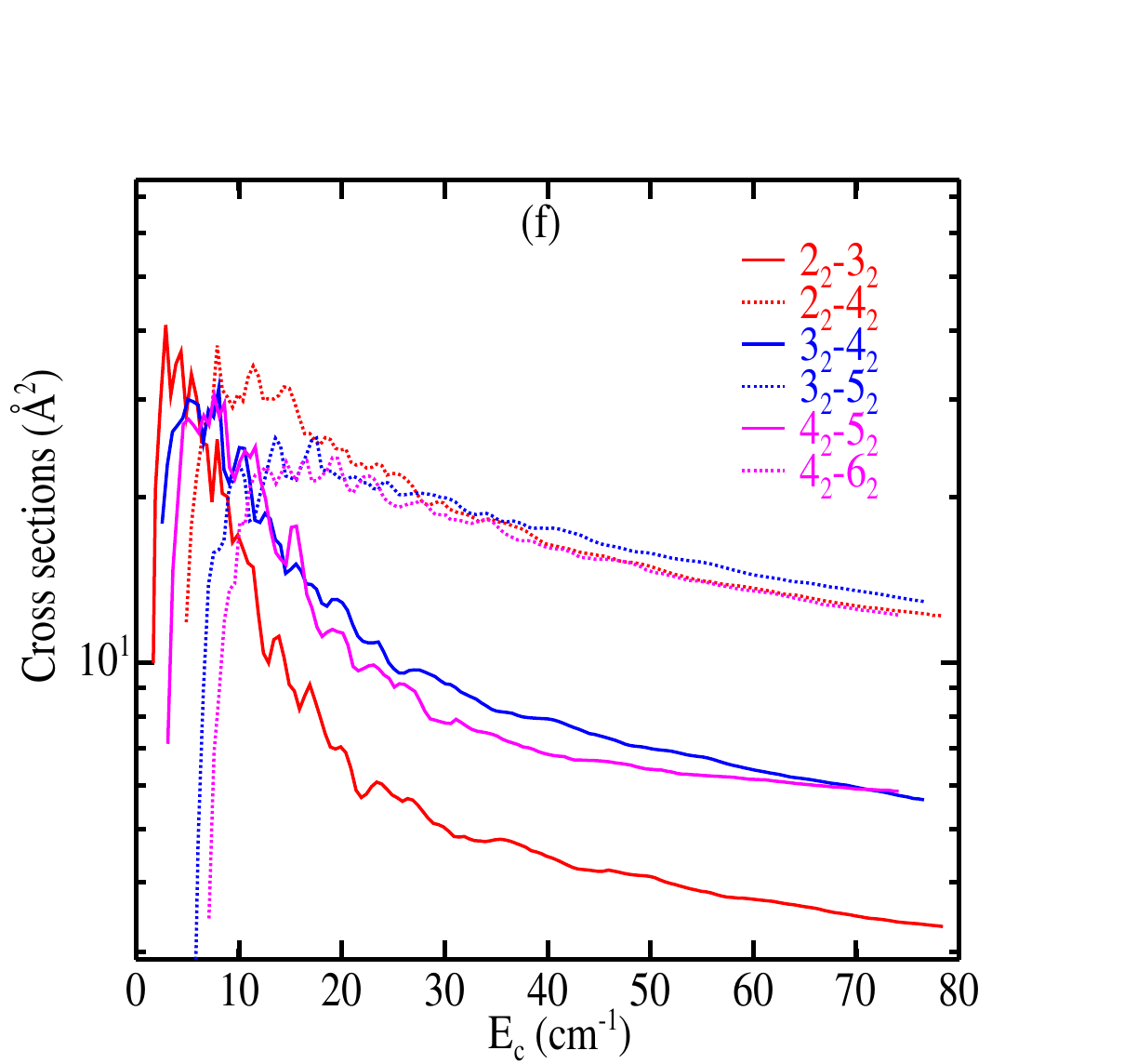}
\end{minipage}%
\caption{Kinetic energy dependence of the rotational excitation cross sections $j_k \rightarrow j'_{k'}$ of $para-$CH$_3$NC-He (left panels) and $para-$CH$_3$CN-He (right panels) in collision with He. (a), (b): transitions from the ground rotational state ($1_1$) to various $j'_k$ ; (c), (d): comparison of $\Delta j=1$ and $\Delta j=2$ transitions while $k=k'=1$; (e), (f): comparison of $\Delta j=1$ and $\Delta j=2$ transitions while $k=k'=2$.} %; Fourth row, comparison of $\Delta k=0$ and $\Delta k=1$ transitions while $\Delta j = 1$.}
\label{XCS-E}
\end{figure}

\begin{figure}[!ht]
%    \textbf{CH$_3$NC-He \hspace{5cm} CH$_3$CN-He}
\centering
\begin{minipage}{.5\textwidth}
\includegraphics[width=0.8\textwidth]{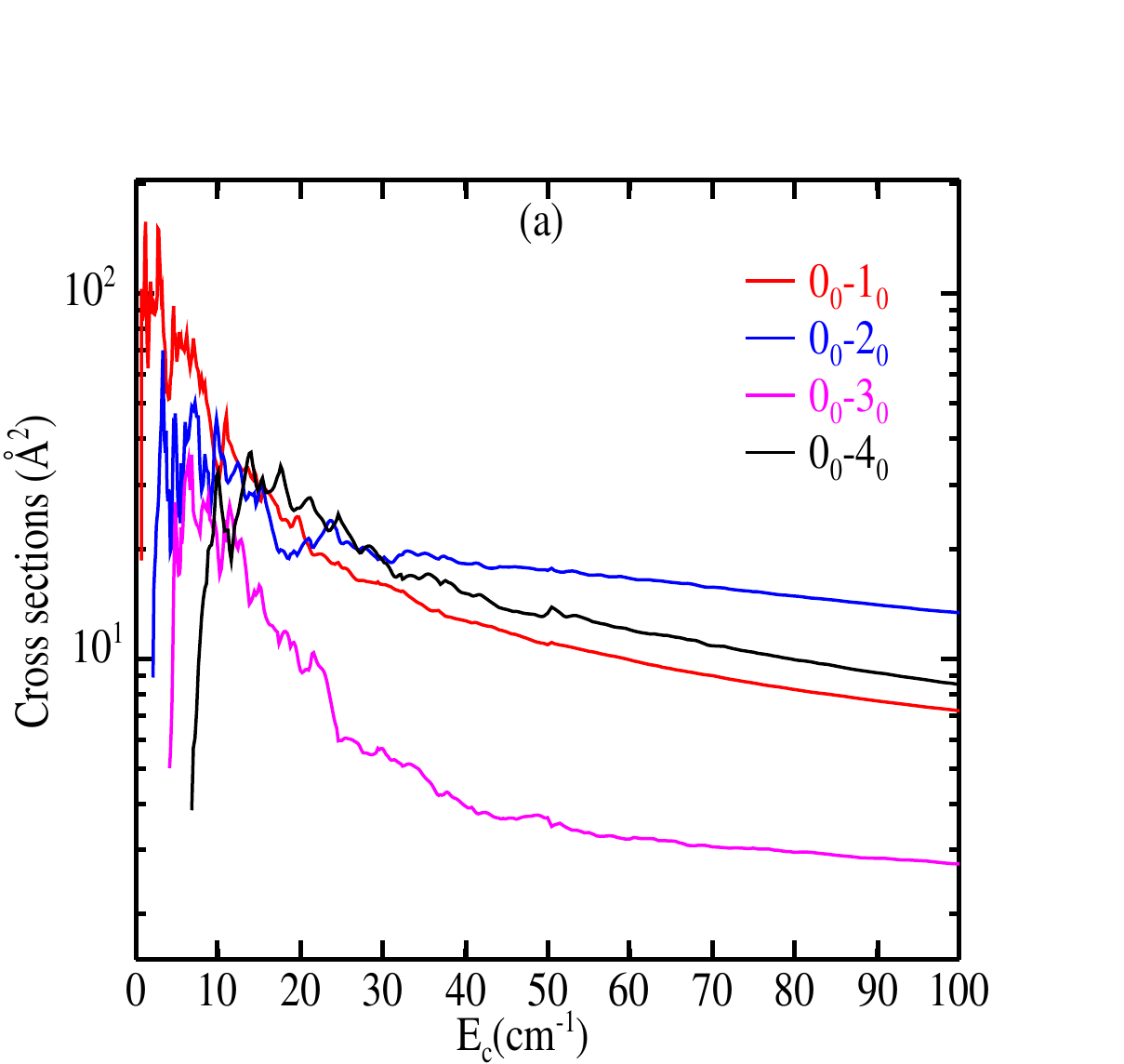}\label{ref:111}
\includegraphics[width=0.8\textwidth]{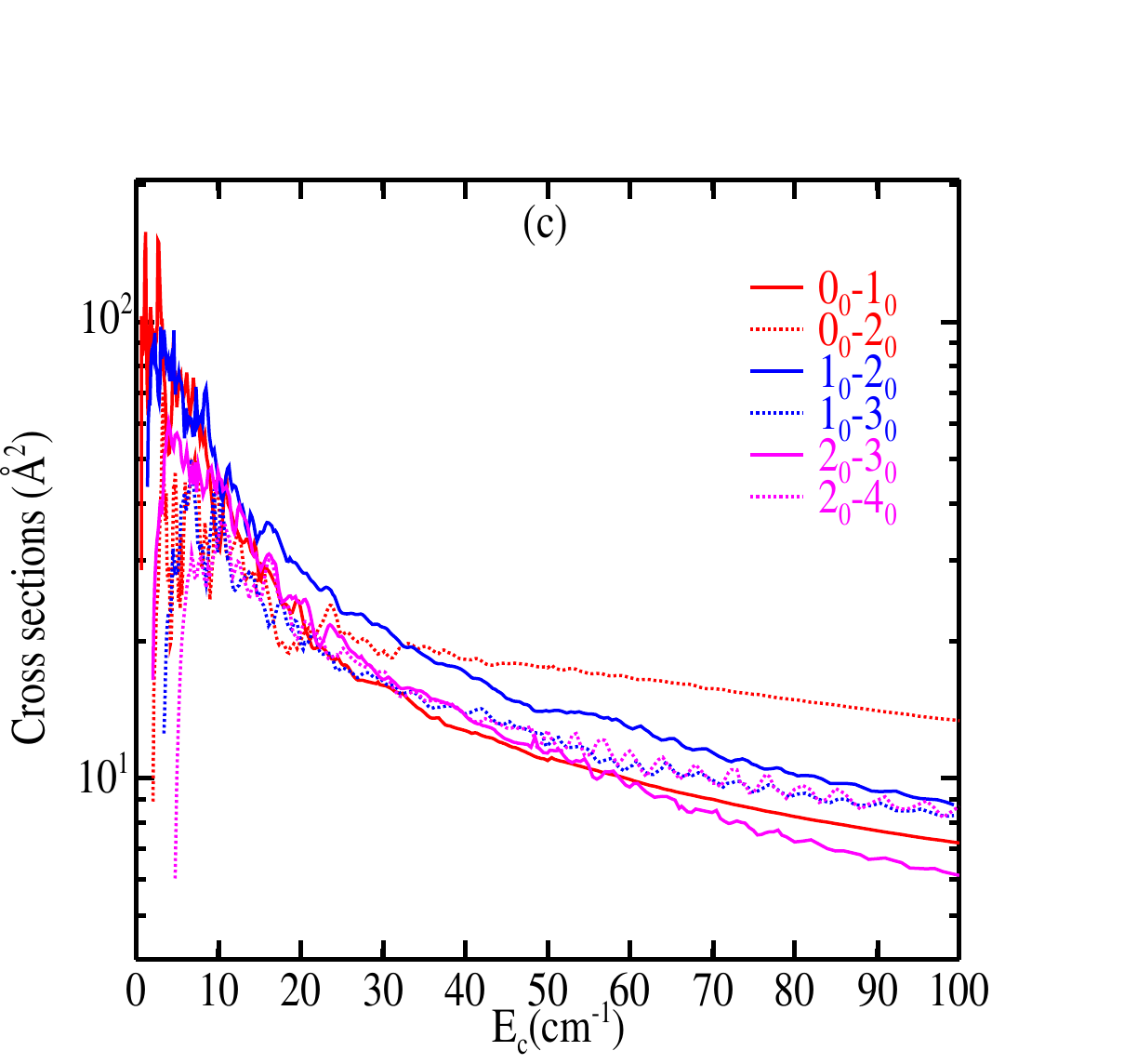}\label{ref:111}
\includegraphics[width=0.8\textwidth]{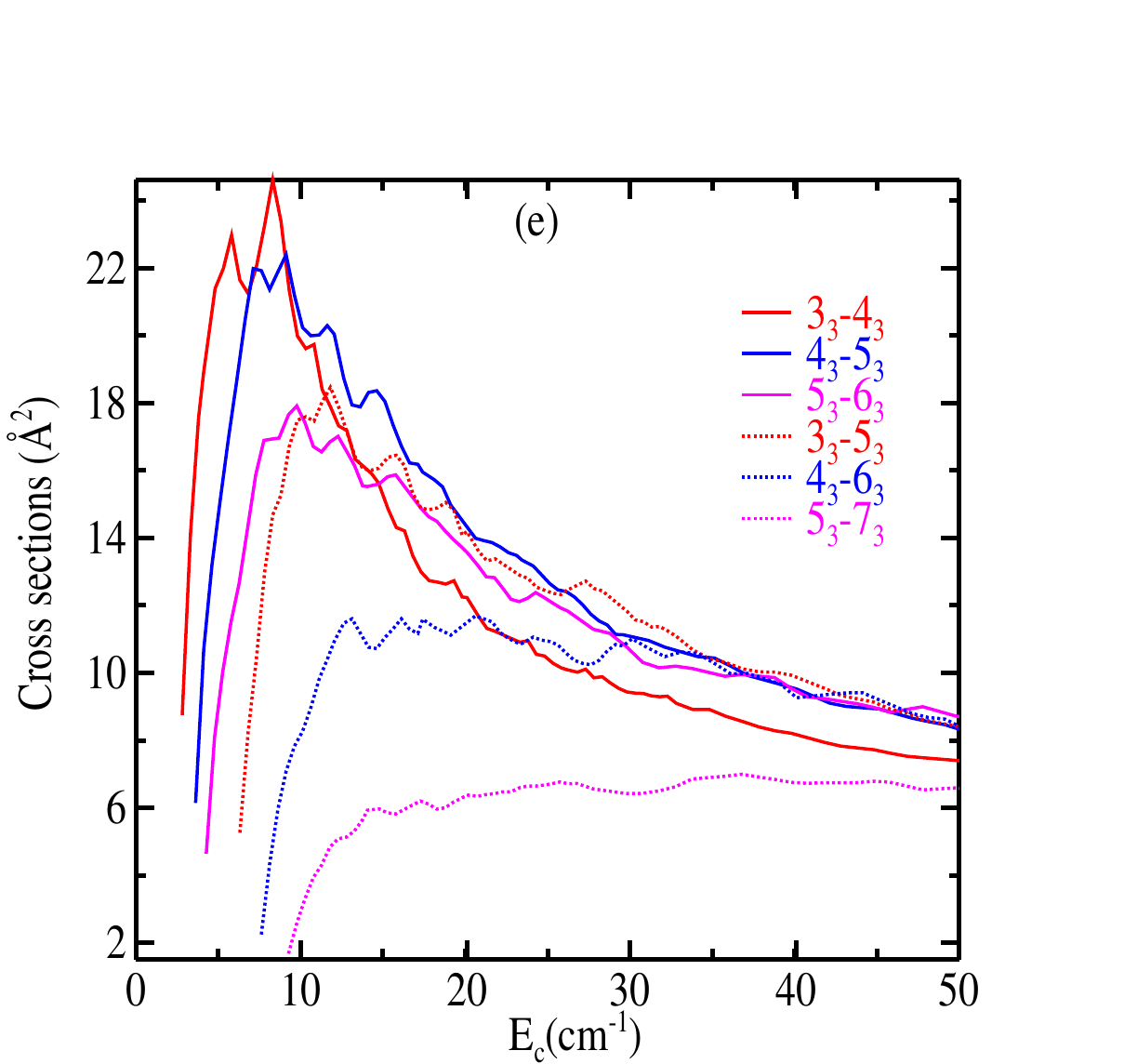}\label{ref:111}
\end{minipage}%
\begin{minipage}{.5\textwidth}
\includegraphics[width=0.8\textwidth]{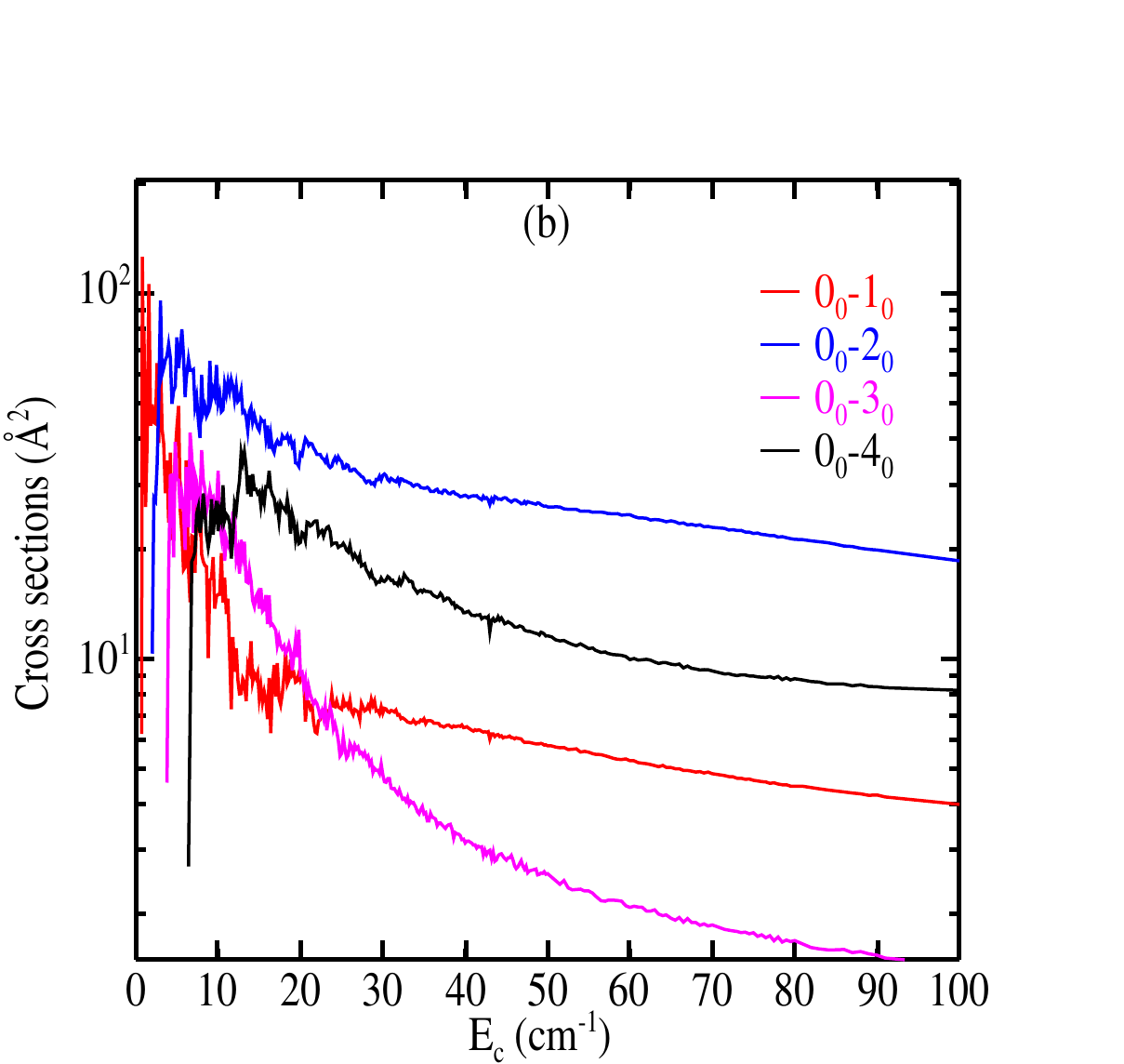}\label{ref:111}
\includegraphics[width=0.8\textwidth]{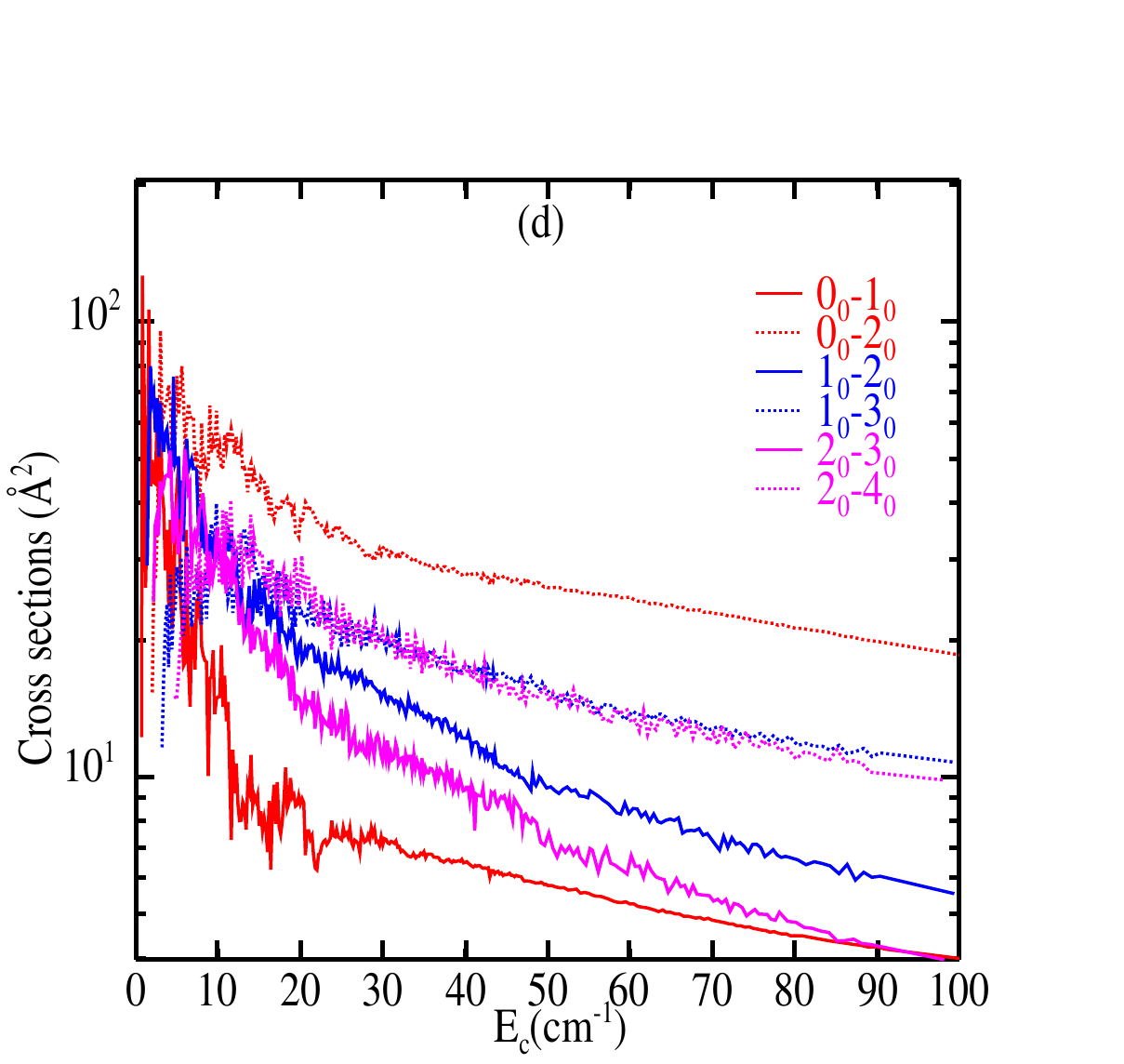}\label{ref:111}
\includegraphics[width=0.8\textwidth]{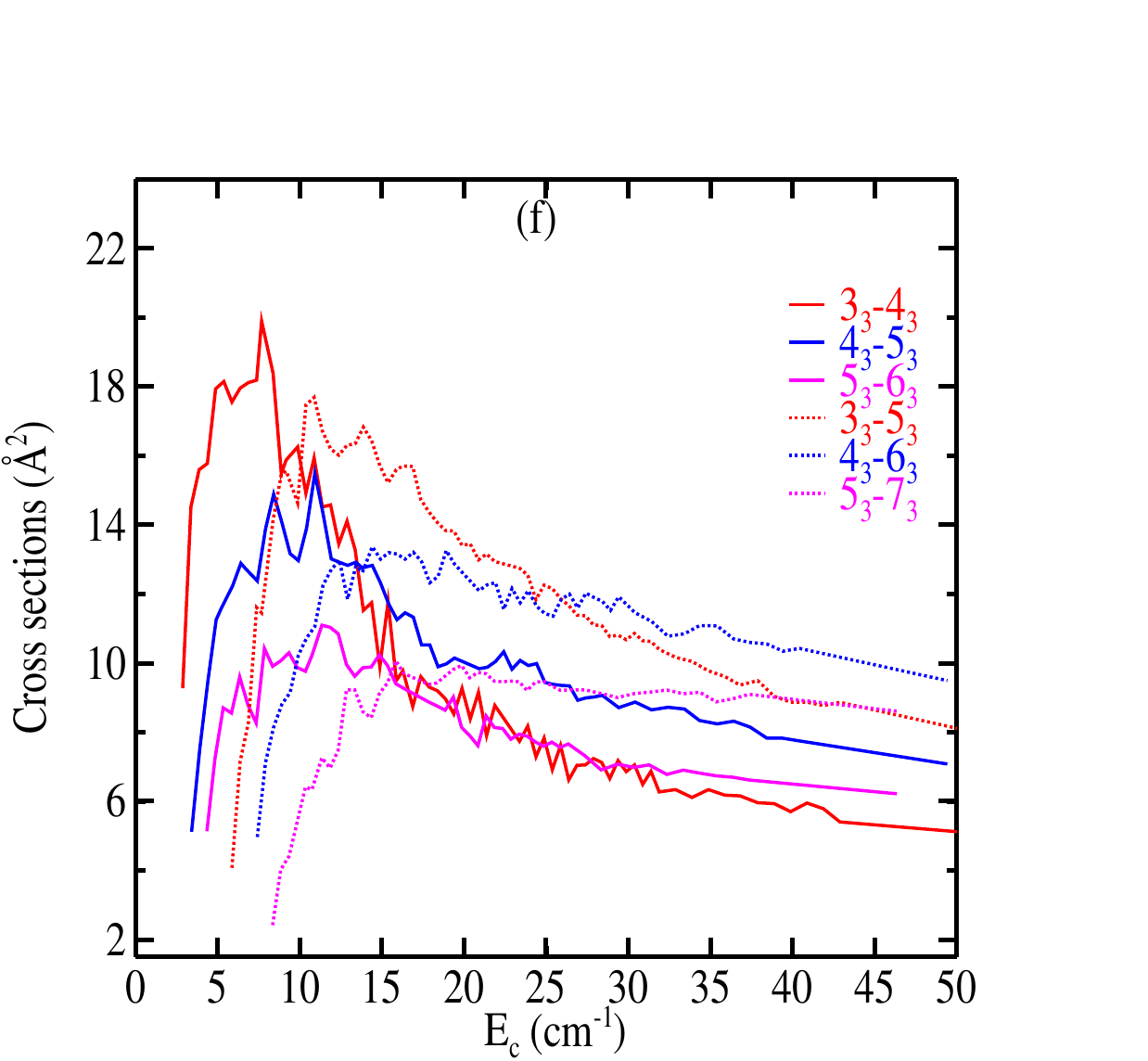}\label{ref:111}
\end{minipage}%
\caption{Kinetic energy dependence of the rotational excitation cross sections $j_k \rightarrow j'_{k'}$ of $ortho-$CH$_3$NC-He (left panels) and $para-$CH$_3$CN-He (right panels) in collision with He. (a), (b): transitions from the ground rotational state ($0_0$) to various $j'_k$; (c), (d): comparison of $\Delta j=1$ and $\Delta j=2$ transitions while $k=k'=0$; (e), (f): comparison of $\Delta j=1$ and $\Delta j=2$ transitions while $k=k'=3$.}
\label{XCS-A}
\end{figure}

Figures~\ref{XCS-E} and ~\ref{XCS-A} present the plots of collisional excitation cross sections of $para-$CH$_3$NC/CN-He and $ortho-$CH$_3$NC/CN-He, as a function of the kinetic energy for transitions from the
ground state, namely 0$_0$ for $ortho$ species and 1$_1$ for $para$ species to levels $j_k$ (Panels a and b),
and for dipolar ($\Delta j=1$) and quadripolar ($\Delta j=2$) transitions with $k=0,1,2,3$ (Panels c,d,e and f).
We note the presence of both Feshbach and shape resonances for kinetic energies below 60 cm$^{-1}$.
Such features are expected, since the global minima of the PESs of the CH$_3$NC-He and CH$_3$CN-He complexes are found to be slightly less than 60 cm$^{-1}$. These resonances arise from the tunnelling of the projectile through the centrifugal energy
barrier (shape resonances) or to a temporary trapping of the projectile by the potential leads to CH$_3$NC-He and CH$_3$CN-He complexes quasi-bound states (Feshbach resonances). The character of the individual resonances can be analysed by inspecting the wave functions \cite{ma2015resonances}.
The resonances disappear and the cross sections becomes smooth and continue to decrease for energies $E_{tot} \geq$ 60 cm$^{-1}$.
Comparing the cross sections for CH$_3$NC-He (left panels) and CH$_3$CN-He (right panels), we can see that the relative order of magnitudes of the CH$_3$NC-He cross sections are larger than those for CH$_3$CN-He, which can be explained by the larger well depth of the CH$_3$NC-He PES compared to that of the CH$_3$CN-He PES.

In addition, we note also significant differences between the CH$_3$CN-He and CH$_3$NC-He cross sections. Figures~\ref{XCS-E} and ~\ref{XCS-A} reveal the existence of a strong even $\Delta j$ propensity
rule for almost all energies for CH$_3$CN-He. The largest cross sections are
found for the quadripolar transitions 0$_0-2_0$, 1$_0-3_0$ and 2$_0-4_0$ for $ortho-$CH$_3$CN-He and for the transitions 1$_1-3_1$, 2$_1-4_1$ and 3$_1-5_1$ for $para-$CH$_3$CN-He. 
On the other hand, for its isomer CH$_3$NC-He, the predominance of cross sections is associated to transitions with |$\Delta j|=1$ for $E_c \leq$ 40 cm$^{-1}$. It should be noted that this propensity rule changes with increasing energy, and transitions |$\Delta j|=2$ become favored for $E_c \geq$ 40 cm$^{-1}$.  
The  propensity rules of the collisional cross sections can be understood by examining the shape of the PES : the higher symmetry of the CH$_3$CN-He PES will favour transitions with
even |$\Delta j|$ whereas the anisotropic PES of CH$_3$NC-He favors transitions with odd |$\Delta j|$ \cite{alexander1982propensity}. In addition, this propensity rule can be explained by interpreting
the expansion coefficients $V_{lm}(R)$ presented in figure~\ref{coeff}. For example, inelastic cross sections involving
|$\Delta j|=1$ are caused by the $V_{10}$ term, which is the dominant term for CH$_3$NC-He, so that the cross section corresponding to the 1$_0$-0$_0$ transition is larger than for other transitions. However, for CH$_3$CN-He, the dominant term is $V_{20}$, a term responsible for transitions involving |$\Delta j|=2$, hence, the largest cross sections are obtained for even $\Delta j$ transitions.
Such results were already observed for other systems : propensity rules which favor |$\Delta j|=2$ were also found for X-CN systems such as SiH$_3$CN \cite{naouai2021inelastic}, HCN \cite{sarrasin2010rotational}, CCN \cite{chefai2018collisional} 
and which favor |$\Delta j|=1$ for X-NC complex such as HMgNC \cite{amor2021rotational}, HNC \cite{sarrasin2010rotational}, NaNC \cite{gharbi2021sodium} as well as MgNC and AlNC \cite{hernandez2013cyanide}.
Recently, collisions of HCN-He and HNC-He were investigated experimentally with chirped pulse in uniform supersonic flow technique by \citet{hays2022collisional}.
Good agreement between the scattering calculations and experimental measurements was found, and similar effects of propensity rules were observed for collisions
of HCN and HNC with He atoms.

Finally, we note that for $para$ states collisional cross sections for transitions with |$\Delta j| =1$ and |$\Delta k|=0$ are strongly dominant over transitions associated with |$\Delta j|$=1 and |$\Delta k|=1$ (not shown). In addition, resonances almost do not exist for cross sections involving levels with |$\Delta k|=1$, due to the large threshold energy value. 

It should be noted that no rate coefficients for CH$_3$NC excitation by collision with helium are available in the literature, and the interpretation of astrophysical observation is usually realised by using the CH$_3$CN-He collisional rates, since the dipole moment of CH$_3$NC and its isomers are almost identical ($\mu$=3.92 D for CH$_3$CN and $\mu$=3.89 D for CH$_3$NC) and the energies of the rotational levels are almost equal.
However, the comparison of the corresponding cross sections of both isomers displays large differences that will be reflected on the magnitude of collisional rate coefficients. 
We conclude that the use of CH$_3$CN rate coefficients in the interpretation of CH$_3$NC observations  may be dangerous for the astrophysical applications, since the use of CH$_3$CN collisional
rate coefficients will probably significantly modify the excitation of the CH$_3$NC molecule. This will be explored in future research.

\section{Conclusions}\label{sec:conclusion}
In this paper, we have calculated accurate, state-of-the-art 3D-PESs corresponding to the interaction of rigid CH$_3$CN and CH$_3$NC molecules with an He atom using the CCSD(T)-F12a/aug-cc-pVTZ level of theory to study the rotational excitation induced by molecule-atom collisions. The potential well depths are moderate, with a global minimum of about 58 cm$^{-1}$ for CH$_3$NC-He and 55 cm$^{-1}$ for CH$_3$CN-He, allowing a high quality study of the dynamics.
Using these PESs, we have computed cross sections for total energies ranging from the threshold up to 100 cm$^{-1}$ for both isomers. Our calculations show the dominance of $\vert\Delta j \vert  =1$ transitions for CH$_3$NC-He collisions and a dominance of $\vert\Delta j \vert =2$  transitions for CH$_3$CN excitation by He atoms. 

CH$_3$CN is one of the most ubiquitous detected molecule in astrophysical clouds, where it acts as practical gas thermometers of the interstellar cloud. Its isomer CH$_3$NC is the less stable isomer of CH$_3$CN, and astrophysicists have often used CH$_3$CN in order to understand the abundance of interstellar CH$_3$NC. As demonstrated in this paper, cross sections computed based on an appropriate PESs and with quantum-mechanical techniques are very different between both isomers. We thus expect this to impact significantly the abundance of the CH$_3$NC isomer, when non-local thermodynamic equilibrium conditions prevail. The high energy scattering as well as the corresponding rate coefficients and their impact on the interpretation of observations of methyl (iso)-cyanide will be published in a forthcoming paper.

\section*{Acknowledgments}
J.L. acknowledges support from KU Leuven through Grant No. 19-00313.


\providecommand{\latin}[1]{#1}
\providecommand*\mcitethebibliography{\thebibliography}
\csname @ifundefined\endcsname{endmcitethebibliography}
  {\let\endmcitethebibliography\endthebibliography}{}

\end{document}